\documentclass[runningheads]{llncs}
\usepackage[T1]{fontenc}
\usepackage{graphicx}
\usepackage{amsmath,amssymb}
\usepackage{algorithm}
\usepackage{algpseudocode}
\usepackage{placeins}
\usepackage[hidelinks]{hyperref}
\usepackage{booktabs}

\begin{document}
\title{Overlapping Network Community Detection Using Sparse Backbones}
\titlerunning{Overlapping Network Community Detection Using Sparse Backbones}
\author{Zihe Zhou\inst{1}\orcidID{0009-0009-9000-7436} \and
Samin Aref\inst{1}\orcidID{0000-0002-5870-9253}}
\authorrunning{Z. Zhou and S. Aref}
\institute{Department of Mechanical \& Industrial Engineering, University of Toronto,\\ 5 King's College Rd, Toronto, ON M5S 3G8, Canada\\	
\email{zhchen.zhou@mail.utoronto.ca, s.aref@utoronto.ca}}
\maketitle

\begin{abstract}
Community structures are common in real networks, and extracting them provides valuable insight in applications ranging from drug discovery to market segmentation. Overlapping community detection (OCD) is the task of clustering networked data in which nodes may belong to multiple clusters. Existing OCD algorithms often struggle to achieve a suitable balance between detection quality and scalability. We, therefore, propose \textsc{Highway}, a scalable OCD algorithm that exploits the sparse backbone of the input network to perform efficient community inference. We used 728 Lancichinetti-Fortunato-Radicchi benchmark networks to compare \textsc{Highway} and its ablated version against 10 existing OCD algorithms. 
Our results, based on five performance measures, demonstrate a competitive performance for \textsc{Highway}. 
It ranks first in overlapping normalized mutual information with a 6.9\% improvement over the strongest baseline.
It also ranks second in all the other four performance measures. 
These comparative results suggest that \textsc{Highway} coupled with its backbone procedure offers a suitable accuracy-efficiency trade-off.
The \textsc{Highway} algorithm is open-source and available as part of the \href{https://cdlib.readthedocs.io}{CDlib} library.\\ 

This is a post-peer-review accepted manuscript from the Proceedings of the 18th International Conference on Advances in Social Networks Analysis and Mining. The publisher authenticated version in LNCS (version of record) and full citation details are available on Springer's website.


\end{abstract}

\section{Introduction}
\label{sec:introduction}

Community detection is a fundamental problem in network science and has been extensively studied over the past decades \cite{xie_overlapping_2013}. In many real-world networks, nodes can belong to multiple communities, such as social networks \cite{cazabet_detection_2010,goldberg_finding_2010} and biological networks \cite{palla_uncovering_2005}. 
Modeling this multi-membership structure is essential for studying networked systems through accurately identifying their underlying community structure. 
Despite substantial progress in OCD methods \cite{ribeiro_santiago_comparing_2024}, many approaches remain computationally expensive in large or dense networks.
A major reason is that inference is usually performed on the full graph, where weakly informative edges can increase computational cost and obscure community signals. 
In contrast, prior studies have shown that network backbone structures preserve essential connectivity patterns while providing a compact representation of the network \cite{batagelj_om_2003,coscia_network_2017,karrer_robustness_2008}. 
Building on this insight, we propose an algorithm for OCD that operates on a network backbone and is designed to be more scalable than existing methods. Unlike existing approaches that perform inference on the full graph, our method explicitly identifies structurally informative edges that strengthen community signal propagation to form a backbone graph. After selecting nodes as signal sources, we perform membership inference on this reduced structure. This design substantially reduces redundant dependency updates while preserving essential community information, enabling high scalability and competitive performance compared with existing OCD methods.

This paper is structured as follows. After providing the background and stating the problem in Section \ref{sec:problem}, we describe the proposed method in four subsections within Section \ref{sec:method}. Section \ref{sec:results} provides experimental setup and results comparing the proposed method against existing baselines. Finally, Section \ref{sec:conclusion} discusses results and concludes the study.

\section{Background and Problem Statement}
\label{sec:problem}

Existing OCD methods fall into four main paradigms: label propagation algorithms \cite{xie_slpa_2011,gregory_finding_2010}, clique percolation methods \cite{ribeiro_santiago_comparing_2024,palla_k-clique_2008}, optimization-based models \cite{yang_overlapping_2013}, and local expansion methods \cite{baumes_efficient_2005,hollocou_improving_2016,hollocou_multiple_2018}. These families differ in modeling assumptions and complexity, leading to distinct trade-offs between scalability and detection performance. Several detailed reviews of OCD methods are available in the literature \cite{xie_overlapping_2013,gupta_review_2022}.

The OCD problem can be stated as follows. Given a network $G = (V, E)$, with size $m = |E|$ and order $n=|V|$, let $A \in \{0, 1\}^{n \times n}$ denote its adjacency matrix, where $A_{uv} = 1$ if $(u, v) \in E$ and $A_{uv} = 0$ otherwise. The objective of OCD is to identify a collection of possibly overlapping communities $\mathcal{C} = \{C_1, C_2, \dots, C_K\}$ \cite{lancichinetti_community_2009}, where $K$ is the number of communities either specified by the user or returned by the detection method. 

The output of OCD can be represented by a membership matrix $\alpha \in \mathbb{R}^{n \times K}$ \cite{xie_overlapping_2013} to quantify the extent of belonging node $i$ has with cluster $C_j$. The belonging factors are normalized so that they add up to 1 for each node \cite{gregory_fuzzy_2011}: $0 \leq \alpha_{ij} \leq 1, \sum_{j=1}^K \alpha_{ij} = 1$. Row $i$ of the membership matrix is a membership vector which represents the overlapping assignment of node $i$ to the $K$ communities.

\section{Proposed Method}
\label{sec:method}

In this section, we describe the \textsc{Highway} algorithm. Our central design assumption is that primary community signals are propagated through a limited set of structurally informative edges. Removing redundant edges via backbone construction may reveal clearer community boundaries while reducing computational overhead.
Based on this idea, \textsc{Highway} extracts a sparse network backbone that preserves key edges for community information flow and then propagates anchor memberships on this backbone to detect overlapping communities. The algorithm consists of four steps: backbone construction, anchor-based initialization, neighbor-only propagation, and anchor-preserving pattern calibration.

\subsection{Backbone Construction}

The first step of \textsc{Highway} extracts a sparse backbone $H = (V, E_H)$ from the original graph $G = (V, E)$.
Each edge is assigned a hybrid importance score based on a modularity-inspired term and a local neighborhood-overlap term.

Let $d_u$ and $d_v$ denote the degrees of nodes $u$ and $v$, respectively. 
For each edge $(u,v)\in E$, we define an edgewise modularity-based score as $s_{\mathrm{mod}}(u,v) = 1 - ({d_u d_v} /{2m})$. 
This term is derived from the overlapping modularity matrix contribution $A_{uv} - {d_u d_v}/{2m}$ \cite{nicosia_extending_2009}, specialized to the observed edges where $A_{uv}=1$. 
It penalizes edges between high-degree nodes whose connections may be explained by high-degree effect rather than community structure.

In addition, we define a Jaccard neighborhood-overlap score as $s_{\mathrm{jac}}(u,v) = ({|\mathcal{N}(u)\cap \mathcal{N}(v)|})/({|\mathcal{N}(u)\cup \mathcal{N}(v)|})$ where $\mathcal{N}(u)$ and $\mathcal{N}(v)$ are neighbor sets of $u$ and $v$ \cite{jaccard1901etude}. 
This term favors edges embedded in coherent regions with strong neighborhood overlap, which tend to lie within stable community structures. 

The hybrid importance score is $s(u,v) = \omega \, s_{\mathrm{mod}} (u,v) + (1-\omega)\, s_{\mathrm{jac}}(u,v)$, where $\omega \in [0,1]$ controls the trade-off between the global degree-corrected importance and local structural similarity. 
For each node $u$, we retain the top-$r_h$ incident edges according to the hybrid score $s(u, v)$. Formally, we define the local retention set as $\mathcal{N}_{H}(u) = \operatorname{Top}_{r_h} \left\{ v \in \mathcal{N}(u)\mid s(u, v) \right\}$, where $\operatorname{Top}_{r_h}$ returns the $r_h$ neighbors of $u$ with the highest edge scores. 
The backbone edge set $E_H$ is obtained by symmetrizing these local selections as $E_H = \{(u,v)\in E : v\in \mathcal{N}_{H}(u) \ \text{or} \ u\in \mathcal{N}_{H}(v)\}$. The resulting backbone $H = (V, E_H)$ is a spanning subgraph of $G$ that preserves only high-scoring edges of each node.

\subsection{Anchor-based Signal Initialization}

After constructing the backbone graph $H$, \textsc{Highway} selects anchor nodes in the full graph $G$ to initialize membership propagation.
Let $A = \{a_1, a_2, \dots, a_q\}$ denote the anchor set. Each anchor $a_c$ corresponds to an anchor index $c$, which is an integer. For each node $v$, let $\alpha_v(c)$ denote the membership entry of node $v$ with respect to the anchor index $c$. Before propagation, we initialize each anchor by setting $\alpha_{a_c}(c) = 1$, while all other membership entries are initialized as zero. 

To avoid redundant anchors in the same local region, \textsc{Highway} uses a greedy degree-cover strategy on the full graph $G$.
Nodes are sorted in decreasing order of degree. A node is selected as an anchor only if it has not been \textit{covered}. 
Once selected, the node and its neighbors in $G$ are marked as \textit{covered}.
This strategy favors structurally influential nodes while keeping them spatially distributed. We select anchors in the full graph because the backbone graph $H$ is already sparsified while the full graph provides stable structural coverage.

\subsection{Neighbor-only Propagation}

Given the backbone graph $H$ and the anchor set $A$, \textsc{Highway} propagates the anchor membership throughout the backbone.
Unlike traditional propagation methods \cite{xie_overlapping_2013}, \textsc{Highway} follows the \emph{neighbor-only principle}: a node cannot reinforce its own previous membership. Its updated membership is determined solely by its backbone neighbors.
This design reduces self-reinforcement bias and requires that each active anchor index be supported by the backbone structure.

At each propagation iteration, the anchor membership of index $c$ is updated by $\alpha_v(c) = \sum_{u \in \mathcal{N}_H(v)} w_{uv}\alpha_u(c)$ where $w_{uv} = {1}/{\sqrt{d_u d_v}}$. Specifically, $\mathcal{N}_H(v)$ denotes the backbone neighbors of $v$ and $d_u, d_v$ denote the degrees of the nodes $u$ and $v$ in the backbone $H$. This weighting mechanism follows the symmetric degree normalization commonly used in local spectral propagation \cite{li2018local}. 
After each update, only the top-$r_p$ strongest anchor indices will be retained and the membership vector is normalized. The tunable parameter $r_p$ controls the number of retained anchor indices per node. In implementation, if neighbors contribute an empty membership during the update, the original membership will be retained.

For $T$ propagation iterations, time complexity is $O(T r_p |E_H|)$ and memory requirement is $O(n r_p)$. Since $|E_H| \ll |E|$ for small retention values, e.g., $r_h \leq 3$, anchor membership propagation is much cheaper than full graph propagation.

\subsection{Anchor-Preserving Pattern Calibration}

Finally, \textsc{Highway} calibrates the propagated anchor memberships into overlapping community assignments. After propagation, each node $v$ has a sparse membership vector $\alpha^{\mathrm{prop}}_v$.
We define its propagated pattern as the set of retained anchor indices, where $S_v=\{c:\alpha^{\mathrm{prop}}_v(c)>0\}$. 
Nodes with the same pattern are treated as intermediate structural units for calibrating memberships.

Let $\mathcal{P}$ denote the set of distinct patterns, and let $V_P=\{v\in V:S_v=P\}$ be the node set associated with pattern $P$.
\textsc{Highway} estimates the structural reliability of the pattern $P$ using its internal edges $e_{\mathrm{in}}(P)$ and external connections $e_{\mathrm{out}}(P)=\sum_{Q\in\mathcal{N}_{\mathcal{P}}(P)}e_{\mathrm{out}}(P,Q)$. Specifically, it calculates an internal ratio, $\rho_{\mathrm{self}}(P)=2e_{\mathrm{in}}(P)/(2e_{\mathrm{in}}(P)+e_{\mathrm{out}}(P))$.
In addition, \textsc{Highway} calculates the normalized entropy on the distribution of edges from $P$ to its neighboring patterns as $H_{\mathrm{out}}(P) = -({\sum_{Q\in\mathcal{N}_{\mathcal{P}}(P)}p_{P,Q}\log p_{P,Q}})/({\log |\mathcal{N}_{\mathcal{P}}(P)|})$ where $p_{P,Q}=e_{\mathrm{out}}(P,Q)/\\e_{\mathrm{out}}(P)$ and $\mathcal{N}_{\mathcal{P}}(P) = \{Q\in\mathcal{P}:Q\neq P,\ e_{\mathrm{out}}(P,Q)\geq 1\}$. 
We set $H_{\mathrm{out}}(P)=0$ when $e_{\mathrm{out}}(P)=0$ or $|\mathcal{N}_{\mathcal{P}}(P)|\leq 1$. Combining them together, \textsc{Highway} defines the pattern confidence score as
\begin{equation}
\label{eq:anchor_preserving_pattern_confidence}
\gamma(P)
=
\mathrm{clip}_{[\gamma_{\min},\gamma_{\max}]}
\left(
\frac{
w_{\mathrm{self}}\rho_{\mathrm{self}}(P)
+
w_{\mathrm{ent}}\bigl(1-H_{\mathrm{out}}(P)\bigr)
}{
w_{\mathrm{self}}+w_{\mathrm{ent}}
}
\right),
\end{equation}
where $w_{\mathrm{self}},w_{\mathrm{ent}}\geq 0$ are tunable weights, and $\gamma_{\min}$ and $\gamma_{\max}$ define the lower and upper bounds of the confidence score. In addition, \textsc{Highway} computes a neighboring pattern consistency score on the backbone graph $H$ as $s_v=|\{u\in N_H(v):P_u=P_v\}|/|N_H(v)|$. A high value of $s_v$ indicates that node $v$ is likely located in a reliable core region. Therefore, the calibration strength is defined as $\lambda_v=\eta\,\gamma(P_v)\,s_v^{\tau}$, where $\eta\in[0,1]$ and $\tau \geq 0$ are two tuning parameters.

For each pattern, \textsc{Highway} computes the mean anchor membership of the nodes in $V_P$ as $q_P(c) = (\sum_{u \in V_P} \alpha^{\mathrm{prop}}_u(c))/|V_P|$. For each node $v$, \textsc{Highway} measures how the backbone neighbors support the anchor index $c$ as $q_N(v,c) = ({ \sum_{u\in N_H(v)}\alpha^{\mathrm{prop}}_u(c) })/({ \sum_{c'} \sum_{u\in N_H(v)}\alpha^{\mathrm{prop}}_u(c') })$. Then, \textsc{Highway} combines them as the calibrated membership, $q_v(c)=\beta q_{P_v}(c) +(1-\beta)q_N(v,c)$, where $\beta\in[0,1]$ is a tunable parameter. Therefore, the membership of node $v$ is updated by $\alpha_v(c)=(1-\lambda_v)\alpha^{\mathrm{prop}}_v(c)+\lambda_v q_v(c)$. The resulting community membership is truncated to the top-$r_d$ anchor indices and then normalized. 

\section{Experiments and Results}
\label{sec:results}

We perform a performance evaluation of \textsc{Highway} against 10 existing OCD methods and a full-graph variant of \textsc{Highway}. Specifically, our experiment addresses an essential question: whether backbone inference achieves high scalability while maintaining effective performance.

\subsection{Experimental Setup}

We evaluate \textsc{Highway} on Lancichinetti-Fortunato-Radicchi (LFR) synthetic benchmarks \cite{lancichinetti_benchmark_2008}. 
LFR benchmarks provide explicit control over the mixing parameters $\mu_w$ , which determine the strength of inter-community edges, i.e., the level of noise, thus managing the difficulty of recovering the underlying community structure. 
In the experiment, we generate 728 LFR benchmarks varying by (i) the number of nodes and edges, (ii) the mixing parameters $\mu_w \in \{0.1, 0.2, \dots, 0.7\}$, and (iii) random seeds.

We compare \textsc{Highway} with a set of 10 OCD methods including: \textsc{Walkscan} \cite{hollocou_improving_2016}, \textsc{Lais2} \cite{baumes_efficient_2005}, \textsc{Conga} \cite{gregory_algorithm_2007}, \textsc{Congo} \cite{gregory_fast_2008}, \textsc{Kclique} \cite{palla_k-clique_2008}, \textsc{Demon} \cite{coscia_demon_2012}, \textsc{SLPA} \cite{xie_slpa_2011}, \textsc{COPRA} \cite{gregory_finding_2010}, \textsc{BigClam} \cite{yang_overlapping_2013}, and \textsc{MultiCom} \cite{hollocou_multiple_2018}. These baselines cover several methodological families, including label propagation \cite{xie_slpa_2011,gregory_finding_2010}, generative modeling \cite{yang_overlapping_2013}, clique percolation \cite{palla_k-clique_2008}, and local expansion \cite{baumes_efficient_2005,hollocou_improving_2016,hollocou_multiple_2018}.
As an ablation study, we evaluate a full-graph variant of our method that runs propagation directly on the original graph $G$, and refer to it as \textsc{HighwayFull}.
To assess performance from complementary perspectives, we use the following performance measures: \textit{Fuzzy Rand Index} (FRI) \cite{hullermeier_fuzzy_2014}, \textit{overlapping modularity} ($Q_{ov}$) \cite{nicosia_extending_2009}, the S\o rensen--Dice coefficient \cite{sorensen1948method} (Dice index for short), \textit{composite overlap-aware metric} $F^\ast$ \cite{dewolfe_pragmatic_2026}, and \textit{overlapping normalized mutual information} (ONMI) \cite{lancichinetti_detecting_2009}. For all of these five metrics, higher values indicate better performance. 

\subsection{Experiment Results}

\begin{figure}
\includegraphics[width=0.98\textwidth]{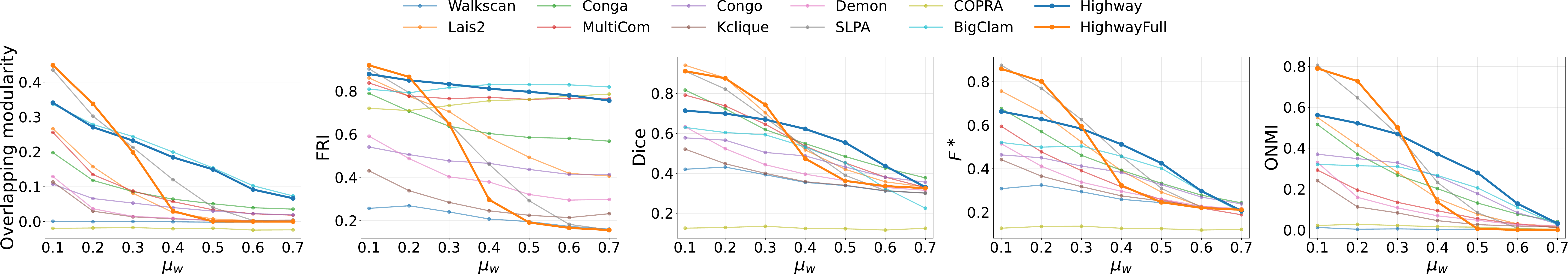}
\caption{Five performance metrics for 12 OCD methods on the LFR benchmarks under increasing mixing parameter $\mu_w$. The high-resolution figure can be magnified on the screen for the details.}
\label{fig:lfr_performance}
\end{figure}

Figure~\ref{fig:lfr_performance} compares five metrics for \textsc{Highway}, its full-graph variant \textsc{HighwayFull}, and 10 algorithms on the LFR benchmark.
Overlapping modularity results in the first panel show that \textsc{Highway} is among the top algorithms across the full range of $\mu_w$ values. It ranks second overall in $Q_{ov}$, within $3.2\%$ of the strongest baseline (\textsc{BigClam}). This indicates that detected assignments continue to preserve the meaningful community structure even as $\mu_w$ increases and the communities become less separable.
Based on FRI, \textsc{Highway} remains competitive throughout the full range of mixing parameter values. Although several methods, such as \textsc{BigClam}, \textsc{SLPA}, and \textsc{COPRA}, are highly competitive based on FRI, \textsc{Highway} stays in the top tier, ranking second and trailing the strongest baseline by only $0.4\%$.
The Dice, $F^\ast$ and ONMI panels further suggest that \textsc{Highway} can recover ground-truth overlapping community structure across different levels of noise. Although \textsc{Highway} is not always the best OCD algorithm at low noise level, its Dice, $F^\ast$, and ONMI performances become competitive as $\mu_w$ increases. \textsc{Highway} ranks second in both Dice and $F^\ast$, within $2.5\%$ and $3.9\%$ of the strongest baseline respectively. In addition, it ranks first in ONMI, exceeding the strongest baseline (\textsc{SLPA}) by $6.9\%$. Specifically, when $\mu_w \geq 0.4$, \textsc{Highway} maintains a stronger agreement on these three supervised metrics with the ground-truth communities than most OCD methods. Beside its superior performance in ONMI, \textsc{Highway} ranks second in all the other four performance measures.
We further conduct an ablation study by comparing \textsc{Highway} and its full-graph variant, \textsc{HighwayFull}, to assess the contribution of backbone construction. Although full-graph propagation performs well under low-noise settings in Figure~\ref{fig:lfr_performance}, the sparse backbone used by \textsc{Highway} improves robustness when the mixing level increases.


\section{Discussion and Conclusions} 
\label{sec:conclusion}

The comparative results presented support the modeling assumption of \textsc{Highway}, that is, selectively disregarding some edges is not only permissible for OCD, but can also be beneficial. 
Compared to ten other methods, \textsc{Highway} ranks first in ONMI and it ranks second in all the other four performance measures. 
The ablation test further suggests that the backbone construction process contributes to detection robustness. When the mixing parameter is small, \textsc{HighwayFull} (w/o backbone) performs well because the community signal is relatively clean. However, as the mixing parameter increases, the full graph contains a smaller proportion of informative edges and relying on them all comes at a cost in performance. \textsc{Highway} (with backbone) maintains stronger performance at medium to high mixing levels. 
Processing the sparse backbone preserves primary community signals while reducing the propagation of noise.
In conclusion, this paper introduced \textsc{Highway}, an OCD method that strikes a balance between performance and efficiency. Avoiding a costly full-graph processing, \textsc{Highway} first extracts a structurally informative backbone and then performs anchor-based propagation and pattern calibration on the backbone subgraph. Experimental LFR results against 10 existing methods showed that \textsc{Highway} achieves the best or the second best performance on all the five evaluation metrics.  

Most existing OCD methods process the full graph, resulting in major time and memory requirements for large networks. Our methodological contributions, therefore, go beyond proposing yet another algorithm. Our study investigated a design paradigm that departs from the traditional full-graph processing for OCD \cite{xie_overlapping_2013}. This paradigm is promising because lightweight inference on a sparse backbone was shown to achieve strong community recovery performance while reducing computational cost. The \textsc{Highway} algorithm is open-source and available as part of the \href{https://cdlib.readthedocs.io}{CDlib} library \cite{rossetti2019cdlib}.

\end{document}